\documentclass[12pt]{iopart}
\usepackage[english,russian]{babel}
\usepackage{wrapfig,graphicx}
\usepackage{amsthm,amssymb}
\usepackage{iopams}
\begin{document}

\title[Dynamics of the Chaplygin ball on a rotating plane]{Dynamics of the Chaplygin ball on a rotating plane}

\author{I.\,A.~Bizyaev$^{1,2}$, A.\,V.~Borisov$^{1,2}$ and I.\,S.~Mamaev$^{1,3}$}
\address{$^1$Moscow Institute of Physics and Technology,
Institutskii per. 9, Dolgoprudnyi, 141700 Russia}
\address{$^2$Udmurt State University,
ul. Universitetskaya 1, Izhevsk, 426034 Russia}
\address{$^3$Izhevsk State Technical University,
ul. Studencheskaya 7, Izhevsk, 426069 Russia}
\ead{bizaev\_90@mail.ru, borisov@rcd.ru and mamaev@rcd.ru}

{\bf Abstract}
This paper addresses the problem of the Chaplygin ball rolling on a horizontal plane which rotates with constant angular
velocity.
In this case, the equations of motion admit area integrals, an integral of squared angular momentum and the Jacobi integral, which is
a generalization of the energy integral, and possess an invariant measure.
After reduction the problem reduces to investigating a three-dimensional  Poincar\'{e} map that preserves phase volume (with density
defined by the invariant measure).
We show that in the general case the system's dynamics is chaotic.

\vspace{2pc}
\noindent{\it Keywords}: Chaplygin ball, nonholonomic mechanics, Poincar\'{e} map,
invariant measure, reduction, first integrals, Jacobi integral, affine constraints.

Mathematics Subject Classification: 37J60

\newpage
\section*{Introduction}
{\bf 1.} Affine (inhomogeneous) constraints in nonholonomic mechanics are represented as
$$
f_{\mu} = \big(\boldsymbol{a}_{\mu}(\boldsymbol{q}), \dot{\boldsymbol{q}}\big) + b_\mu(\boldsymbol{q})=0,
$$
where $\boldsymbol{q}$ and $\boldsymbol{\dot{q}}$ are the generalized coordinates and velocities of the system.
The best understood case is that of constraints homogeneous in velocities (i.\,e., $b_\mu(\boldsymbol{q})=0$). These constraints arise, for example,
when a convex rigid body rolls without slipping or when a rigid body with a fastened wheel pair (knife edge) moves on a fixed
supporting surface (the Chaplygin sleigh).
As was shown already by Hertz, equations of motion preserve the energy integral in the case of homogeneous constraints.
For more detailed treatments of modern methods and problems of nonholonomic mechanics,
see~\cite{KozlovRev,Hierarchy,*2,Acceleration,UMN}.

As is well known, a system with affine constraints ($b_\mu(\boldsymbol{q})\neq0$) does generally not preserve the energy integral.
Nevertheless, an example of constraints in which the system admits
a generalization of the energy integral is provided by the motion of a rigid body on a supporting surface which rotates with
constant angular velocity.
In this case, after transition to a uniformly rotating coordinate system,
terms linear in the velocity are added to the Lagrangian functions,
and the constraints are reduced to homogeneous ones.
For this reason there exists an additional integral, which was
called the Jacobi integral in~\cite{JacobiIntegral} using an analogy with celestial mechanics.
In~\cite{Fas, Fass},  an analogous integral is called ``moving energy''.

{\bf 2.} The best-known example of a system with affine constraints is the rolling motion of a homogeneous ball on a uniformly
rotating plane.
This problem was first considered by S.\,Earnshaw~\cite{Er} before nonholonomic
mechanics grew into a separate discipline~\cite{Historical}.
He showed  that the trajectory of the center of the ball in absolute space is a circle the position of the center of which depends on initial conditions.
Later this problem was considered in the textbooks of E.\,A.~Milne~\cite{Milne}, Yu.\,I.~Neimark and
N.\,A.~Fufaev~\cite{NF}, and E.\,J.~Routh~\cite{Rays}.
The authors of~\cite{bor3} discuss the following interesting demonstration based on this problem and presented in the Franklin
Museum: a homogeneous ball rolling in a straight line gets onto a rotating table and then leaves it along the same straight line.
Theoretical and experimental results on this system, called ANAIS billiard, are discussed in~\cite{ANAIS1, ANAIS2}.
For references to the literature on various generalizations of the Earnshaw problem, see~\cite{JacobiIntegral}.

{\bf 3.} Another well-known integrable, but much more complex system of nonholonomic mechanics is the problem of a dynamically
asymmetric balanced ball~\cite{Chaplygin} ({\it Chaplygin ball}) rolling on a fixed horizontal plane.
In~\cite{Drift}, a detailed qualitative analysis of this system is presented and, in particular, conditions for boundedness and
unboundedness of trajectories of the contact point are found.

This problem admits a number of integrable generalizations obtained by adding a gyrostatic term or a Brun field~\cite{KozlovUSM}.
Another integrable nontrivial generalization is related to the Chaplygin ball rolling over a sphere~\cite{Fedorov, BMF}.
However, if the center of mass of the ball is displaced relative to the geometric center, then
the system becomes nonintegrable~\cite{Kazakov} and exhibits the reversal phenomenon~\cite{Sataev} and strange attractors
typical of rattlebacks~\cite{RattlebackUFN}.
There is no detailed treatment of dynamics in this case, although it is of great interest, especially when it comes to
verifying the nonholonomic model (that is, checking whether it agrees with experiments) and detecting new nontrivial dynamical
phenomena. Particular interest in this problem has been stimulated recently by designs of spherical robots which move
by displacing the center of mass~\cite{Alves, Conticelli}.

{\bf 4.} In this paper, we consider the motion of the Chaplygin ball on a horizontal plane which rotates with constant angular velocity.
An explicit generalization of the energy integral in this case was presented in~\cite{Naranjo},
although this integral can easily be obtained on the basis of the results of~\cite{JacobiIntegral}.
Moreover, this problem admits an area integral, an integral of squared angular momentum, and an invariant measure.
After reduction the problem reduces to investigating a three-dimensional Poincar\'{e} map that preserves phase volume
(with density defined by an invariant measure).
In this paper we show that in the general case the dynamics of the system is chaotic.
Similar three-dimensional maps were considered earlier in relation to chaotic advection in~\cite{Cartwright,Sun}.

Three-dimensional maps that are given by explicit expressions (and not generated by the phase flow) have recently been investigated
in~\cite{James,Dullin} from the viewpoint of bifurcation and chaotic dynamics. Of interest is the generalization of methods which are developed in these studies
for the problem we consider here.
A number of phenomena that are similar to those described in~\cite{James,Dullin} and lead to the appearance of three-dimensional tori
are presented in this paper.
When there is axial symmetry, the problem can be reduced to investigating a two-dimensional area-preserving map where
the dynamics can be analyzed in more detail.

{\bf 5.} We note that Tz\'{e}noff~\cite{Tzenoff,Tzenoff2} considered the problem of a dynamically symmetric body of revolution
rolling on a rotating plane to illustrate a new form (proposed by him) of equations of nonholonomic mechanics.
However, in these papers he failed to obtain any dynamical conclusions or results, and the equations of motion presented in them
are intractable for further analysis. As will be shown below, he also made an incorrect statement concerning integrability of the
case where
the body has a spherical surface.

\newpage
\section{Equations of motion with an inhomogeneous nonholonomic constraint}
\label{sec1}

Consider the problem of an inhomogeneous balanced ball rolling without slipping on a horizontal plane rotating with constant
angular velocity
$\Omega$. Let us choose two coordinate systems:
\begin{itemize}
\item[{--}] $Oxyz$, a fixed coordinate system with a vertical axis $OZ$ coinciding with the
axis of rotation of the plane;
\item[{--}] $Cx_1x_2x_3$, a moving coordinate system with origin $C$ at the center of
the ball and with axes directed along the principal axes of inertia.
\end{itemize}

The position and orientation of the ball are given by coordinates $x$ and $y$ of
the ball's center $C$ on the plane $Oxy$ and by the matrix of rotation of the moving axes relative to
the fixed axes
\begin{equation*}
{\bf Q}=
\left(\eqalign{
\alpha_{1} \quad \alpha_{2} \quad \alpha_{3} \\
\beta_{1} \quad \beta_{2} \quad \beta_{3} \\
\gamma_{1} \quad \gamma_{2} \quad \gamma_{3}
} \right)
\in SO(3).
\end{equation*}

In what follows, unless otherwise specified, all vectors are referred to the moving coordinate system $Ox_1x_2x_3$, and the following notation is
used:
\begin{itemize}
	\item[{--}] $\boldsymbol v=(v_1, v_2, v_3)$~--- the velocity of the ball's center $C$,
	\item[{--}] $\boldsymbol \omega=(\omega_1, \omega_2, \omega_3)$~--- the angular velocity of the ball,
	\item[{--}] $\boldsymbol \alpha=(\alpha_1, \alpha_2, \alpha_3)$, $\boldsymbol \beta=(\beta_1, \beta_2, \beta_3)$, $\boldsymbol \gamma=(\gamma_1, \gamma_2,
	\gamma_3)$~--- the unit vectors of the fixed axes $Oxyz$, with $\boldsymbol \gamma \parallel Oz$.
\end{itemize}

The constraint equations which express the no-slip constraint can be represented in the vector form
\begin{equation}
\label{eq1}
\boldsymbol f=\boldsymbol v-a \boldsymbol \omega\times \boldsymbol \gamma-\Omega \boldsymbol \gamma\times \boldsymbol R=0,
\quad \boldsymbol R=x\boldsymbol\alpha +y\boldsymbol \beta,
\end{equation}
where $\boldsymbol R$~is the radius vector of the ball's center $C$ in the moving axes and $a$~is the radius of the ball.

The kinematic equations governing the evolution of the position and orientation of the ball can be written as
\begin{equation}
\label{eq4}
\dot{\boldsymbol \alpha}=\boldsymbol\alpha\times\boldsymbol\omega, \quad
\dot{\boldsymbol\beta}=\boldsymbol\beta\times\boldsymbol\omega,
\quad \dot{\boldsymbol\gamma}=\boldsymbol\gamma\times\boldsymbol\omega, \\
\dot{x}=(\boldsymbol\alpha, \boldsymbol v), \quad \dot{y}=(\boldsymbol\beta, \boldsymbol v).
\end{equation}

As is well known, dynamical equations of a system are derived from the D'Alembert\,--\,Lagrange principle and have in this case the form~\cite{BorMam2015}:
\begin{eqnarray}
\label{eq3}
\frac{d}{dt}\left(\frac{\partial L}{\partial \boldsymbol v}\right)+\boldsymbol \omega\times \frac{\partial L}{\partial \boldsymbol v}=\frac{\partial L}
{\partial x}\boldsymbol\alpha+\frac{\partial L}{\partial y}\boldsymbol\beta+
\sum\limits_\mu N_\mu\frac{\partial f_\mu}{\partial \boldsymbol v}, \\
\frac{d}{dt}\left(\frac{\partial L}{\partial \boldsymbol \omega}\right)+\boldsymbol \omega\times \frac{\partial L}{\partial \boldsymbol \omega}+
\boldsymbol v\times\frac{\partial L}{\partial \boldsymbol v}= \\
=\frac{\partial L}{\partial \boldsymbol\alpha}\times\boldsymbol\alpha+\frac{\partial L}{\partial \boldsymbol \beta}\times\boldsymbol\beta+
\frac{\partial L}{\partial \boldsymbol\gamma}\times\boldsymbol \gamma+\sum\limits_\mu N_\mu\frac{\partial f_\mu}{\partial \boldsymbol \omega},
\end{eqnarray}
where $f_\mu$~are the components of the vector constraint equation~(\ref{eq1}) and $N_\mu$~are the undetermined multipliers (constraint reactions).

Since the center of mass coincides with the geometric
center, the Lagrangian function coincides with the kinetic energy of the ball:
$$
L=\frac{1}{2}m\boldsymbol v^2+\frac{1}{2}(\boldsymbol\omega, {\bf I}\boldsymbol\omega),
$$
where $m$ is the mass of the ball and ${\bf I}=diag (I_1, I_2, I_3)$~is its tensor of inertia relative to the center of mass.
Substituting into~(\ref{eq3}), we obtain
\begin{equation}
\label{eq2}
m\dot{\boldsymbol v}+m\boldsymbol\omega\times\boldsymbol v=\boldsymbol N, \quad {\bf I}\dot{\boldsymbol\omega}+
\boldsymbol\omega\times{\bf I}\boldsymbol\omega=a\boldsymbol N\times \boldsymbol\gamma.
\end{equation}

From the constraint equation~(\ref{eq1}) we find the relations
\begin{equation}
\label{eq5}
\boldsymbol v=(a\boldsymbol\omega-\Omega\boldsymbol R)\times\boldsymbol\gamma, \quad \dot{\boldsymbol{v}}=(a\dot{\boldsymbol{\omega}}-\Omega\dot{\boldsymbol R})
\times\boldsymbol \gamma+(a\boldsymbol\omega-\Omega\boldsymbol R)\times(\boldsymbol\gamma\times \boldsymbol\omega).
\end{equation}
Using these relations and eliminating the reaction $\boldsymbol N$ and the velocity $\boldsymbol v$ from~(\ref{eq2}) and (\ref{eq4}), we obtain a complete
system, which describes the dynamics, in the form
\begin{eqnarray}
\label{eq6}
{\bf I}\dot{\boldsymbol\omega}+ma^2\boldsymbol\gamma\times(\dot{\boldsymbol\omega}\times \boldsymbol\gamma)=
{\bf I}\boldsymbol\omega\times \boldsymbol\omega-ma\Omega (\boldsymbol \omega, \boldsymbol\gamma)\boldsymbol R\times \boldsymbol\gamma-
ma \Omega(\dot{\boldsymbol R}\times \boldsymbol \gamma)\times \boldsymbol\gamma, \\
\dot{\boldsymbol\alpha}=\boldsymbol\alpha\times\boldsymbol\omega, \quad \dot{\boldsymbol\beta}=
\boldsymbol\beta\times\boldsymbol\omega, \quad \dot{\boldsymbol\gamma}=\boldsymbol\gamma\times\boldsymbol\omega, \\
\dot{x}=-\Omega y+a(\boldsymbol\omega, \boldsymbol\beta), \quad \dot{y}=\Omega x-a(\boldsymbol\omega, \boldsymbol\alpha).
\end{eqnarray}

These equations define the flow on the eight-dimensional phase space $\mathcal{M}^8=SO(3)\times \mathbb{R}^5$.

\newpage
\section{Conservation laws}
\label{sec2}

Let us restrict the Lagrangian function using~(\ref{eq5}) to the constraint (i.\,e., eliminate $\boldsymbol v$):
$$
L^*=\frac{1}{2}(\boldsymbol\omega, {\bf I}\boldsymbol\omega)+\frac{1}{2}ma^2(\boldsymbol\omega\times \boldsymbol\gamma, \boldsymbol\omega\times\boldsymbol\gamma)-
ma\Omega(\boldsymbol\omega, \boldsymbol R)+\frac{1}{2}m\Omega^2\big(x^2+y^2\big),
$$
where for simplicity we have used the identity $\boldsymbol\gamma\times(\boldsymbol R\times \boldsymbol\gamma)=\boldsymbol R$.
Let us define the angular momentum vector of the system
\begin{equation}
\label{eq13}
\boldsymbol M=\frac{\partial L^*}{\partial\boldsymbol\omega}={\bf I}\boldsymbol\omega+ma^2\boldsymbol\gamma\times(\boldsymbol\omega\times\boldsymbol\gamma)-ma\Omega\boldsymbol R.
\end{equation}

From equations~(\ref{eq6}) we find that its evolution is governed by the equation
$$
\dot{\boldsymbol M}=\boldsymbol M\times \boldsymbol\omega.
$$
This equation implies that the vector $\boldsymbol M$ remains constant in the fixed coordinate system $Oxyz$. As a consequence, we find
that the system~(\ref{eq6}) admits three linear first integrals
\begin{equation}
\label{eq7}
F_1=(\boldsymbol M, \boldsymbol\alpha), \quad F_2=(\boldsymbol M, \boldsymbol\beta), \quad F_3=(\boldsymbol M, \boldsymbol\gamma).
\end{equation}

In addition, it was shown in~\cite{JacobiIntegral} that in the system under consideration one can construct an integral similar to
the Jacobi integral in mechanics
\begin{equation}
\label{eq8}
E=\frac{1}{2}\big(\boldsymbol\omega, {\bf I}\boldsymbol\omega+ma^2\boldsymbol\gamma\times (\boldsymbol\omega\times \boldsymbol\gamma)\big)-
\frac{m}{2}\Omega^2\big(x^2+y^2\big).
\end{equation}
This integral was found explicitly in~\cite{Naranjo}.

Another general invariant of the system~(\ref{eq2}) is the invariant measure $\mu=\rho dx dy d^3\boldsymbol\alpha d^3\boldsymbol\beta d^3\boldsymbol\gamma
d^3\boldsymbol\omega$, where density is given by the expression
$$
\rho=\frac{\det({\bf I}+ma^2{\bf E}-ma^2\boldsymbol{\gamma} \otimes \boldsymbol{\gamma})}{\sqrt{1-ma^2\big(\boldsymbol{\gamma},
({\bf I}+ma^2{\bf E})^{-1}\boldsymbol{\gamma}\big)}}.
$$

Equations~(\ref{eq6}) also admit an obvious symmetry field corresponding to
invariance under rotations of the plane of support:
\begin{equation}
\label{eq9}
\hat{\boldsymbol u}_s=-y\frac{\partial}{\partial x}+x\frac{\partial}{\partial y}-\beta_1\frac{\partial}{\partial \alpha_1}+\alpha_1\frac{\partial}{\partial \beta_1}-
\beta_2\frac{\partial}{\partial \alpha_2}+\alpha_2\frac{\partial}{\partial \beta_2}-\beta_3\frac{\partial}{\partial \alpha_3}+\alpha_3\frac{\partial}{\partial \beta_3}.
\end{equation}

\section{Reduction}
\label{sec3}

Eliminating the variable corresponding to the symmetry field~(\ref{eq9}),
we obtain the reduced system describing the evolution of $\boldsymbol\omega$, $\boldsymbol\gamma$ and $\boldsymbol
R$:
\begin{eqnarray}
\label{eq10}
{\bf \tilde{I}}\dot{\boldsymbol \omega}={\bf I}\boldsymbol\omega\times\boldsymbol\omega-
ma\Omega(\boldsymbol\omega, \boldsymbol\gamma)\boldsymbol R\times \boldsymbol\gamma-
ma\Omega(\dot{\boldsymbol R}\times \boldsymbol\gamma)\times\boldsymbol\gamma \\
\dot{\boldsymbol\gamma}=\boldsymbol\gamma\times\boldsymbol\omega, \quad \dot{\boldsymbol R}=
\boldsymbol R\times( \boldsymbol\omega - \Omega\boldsymbol{\gamma})-a\boldsymbol\gamma\times\boldsymbol\omega,
\end{eqnarray}
where ${\bf \tilde{I}} = {\bf I} + ma^2 (\boldsymbol{\gamma}^2{\bf E} - \boldsymbol{\gamma} \otimes \boldsymbol{\gamma})$ is the tensor of inertia relative to
the point of contact.

The system (\ref{eq10}) admits two geometric integrals whose values are fixed:
$$
\boldsymbol\gamma^2=1, \quad (\boldsymbol R, \boldsymbol\gamma)=0.
$$
These relations define in $\mathbb{R}^9$ the seven-dimensional phase space of the reduced system $\mathcal{M}^7\approx TS^2\times
\mathbb{R}^3$.

A special feature of reduction in this case is that the complete set of integrals
turns out to be noninvariant under the action of the symmetry field~(\ref{eq9}):
\begin{equation}
\label{eq11}
\hat{\boldsymbol u}_s(F_3)=0, \quad \hat{\boldsymbol u}_s(E)=0, \\
\hat{\boldsymbol u}_s(F_1)=-F_2, \quad \hat{\boldsymbol u}_s(F_2)=F_1.
\end{equation}

As a result, the system~(\ref{eq10}) admits only three additional first integrals (and not four, as one would expect), which are
invariant under the action~(\ref{eq11}). For example, one can choose:
\begin{eqnarray*}
\tilde{F}_1=F^2_1+F^2_2 + F^2_3=(\boldsymbol M, \boldsymbol M), \quad \tilde{F}_2=F_3=(\boldsymbol M, \boldsymbol\gamma) \\
E=\frac{1}{2}(\boldsymbol M, \boldsymbol\omega) +\frac{1}{2} m a \Omega(\boldsymbol{\omega}, \boldsymbol{R}) -\frac{m\Omega^2}{2}(\boldsymbol R, \boldsymbol R),
\end{eqnarray*}
where $\boldsymbol M$ is given by~(\ref{eq13}).

From the known solutions $\boldsymbol{\omega}(t)$, $\boldsymbol{\gamma}(t)$, $\boldsymbol{R}(t)$ of the reduced system
(\ref{eq10}) at given values of the first integrals $\tilde{F}_1=C^2$, $\tilde{F}_1=M_\gamma$ and at $\boldsymbol{M}\nparallel\boldsymbol{\gamma}$
the orientation of the ball and the motion of the contact point
are defined by the relations
\begin{equation*}
\dot{\boldsymbol{\alpha}}=\frac{\boldsymbol{M}\times\boldsymbol{\gamma}}{\sqrt{C^2 - M_\gamma^2}}, \quad \dot{\boldsymbol{\beta}}=
\frac{\boldsymbol{M} - M_\gamma\boldsymbol{\gamma}}{\sqrt{C^2 - M_\gamma^2}}, \\
x = (\boldsymbol{\alpha}, \boldsymbol{R}), \quad y = (\boldsymbol{\beta}, \boldsymbol{R}),
\end{equation*}
where $\boldsymbol{M}$ is given by (\ref{eq13}).

For the system (\ref{eq10}) to be integrable, there must exist (for suitable system parameters) another pair of tensor
invariants, for example, two integrals, or an integral and a symmetry field.

\section{Poincar\'{e} map}
\label{sec4}

For numerical analysis and illustration of the behavior of the trajectories of (\ref{eq10}), it is convenient to use the method of
Poincar\'{e} section~\cite{Hierarchy, Kuznetsov}. We describe here briefly its construction for the system considered.

We first restrict the system (\ref{eq10}) to the level manifold of the first integrals
$$
\mathcal{M}^5 = \{ (\boldsymbol{\omega}, \boldsymbol{\gamma}, \boldsymbol{R}) | \ \boldsymbol{\gamma}^2=1,
\ (\boldsymbol{R}, \boldsymbol{\gamma})=0, \ \tilde{F}_1(\boldsymbol{z}) = C^2, \
\tilde{F}_2(\boldsymbol{z}) = M_\gamma \}
$$
and obtain a five-dimensional flow with the energy integral $\tilde{E}=E|_{\mathcal{M}^5}$.  To parameterize
$\mathcal{M}^5$, we use the variables $(K, r_1, r_2,  l, g,)$, which we define as
follows:
$$
K = M_3, \quad r_1= R_1\gamma_1 + R_2\gamma_2, \quad r_2 = R_1\gamma_2 - R_2\gamma_1, \\
\tan l = \frac{M_1}{M_2}, \quad \tan g = \frac{ C(M_2\gamma_1  - M_1\gamma_2)}{ M_\gamma K - C^2 \gamma_3},
$$
%
where $l, g\in [0,2\pi)$ are the angle variables.

Next, we fix the level set of the energy integral $\tilde{E} = h$ and thus obtain
a one-parameter family of four-dimensional flows on the manifolds
$
\mathcal{M}^4_h
$,
and as a secant for this flow we choose a manifold given by the relation
$$
g=g_0={\rm const}.
$$
Numerically integrating the system under consideration and finding intersections of
the trajectories with the given section, we obtain a family of three-dimensional maps
$\mathcal{P}^3_{h, g_0}$, which we parameterize by the variables $(l, r_1, r_2)$,
and define the variable $K$ from the energy integral.

Since the system (\ref{eq10}) possesses an invariant measure, the three-dimensional Poincar\'{e} map preserves some volume form.
Below we outline its main properties.

\begin{figure}[!ht]
	\begin{center}
		\includegraphics[totalheight=4.5cm]{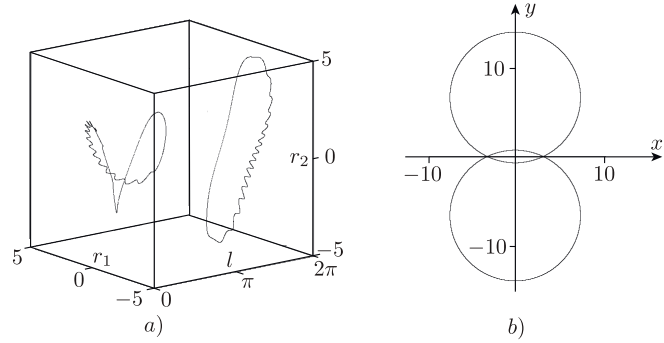}
		\caption{A Poincar\'{e} map for a completely dynamically symmetric ball ($a$) and the trajectory of the
contact point ($b$)
		         for fixed parameters $m=1$, $a=5$, $I_1=I_2=I_3=3$, $H=6$, $h=5$, $G=10$, $\Omega=2$.  }
		\label{fig_Sym}
	\end{center}
\end{figure}

For the homogeneous ball ($I_1=I_2=I_3$) the equations of motion (\ref{eq10}) possess additional symmetry fields corresponding to
invariance of the system under rotation about each axis:
\begin{eqnarray*}
\hat{\boldsymbol u}_1=-\omega_2\frac{\partial}{\partial \omega_3} + \omega_3\frac{\partial}{\partial \omega_2}
-\gamma_2\frac{\partial}{\partial \gamma_3} + \gamma_3\frac{\partial}{\partial \gamma_2}
-R_2\frac{\partial}{\partial R_3} + R_2\frac{\partial}{\partial R_3},\\
\hat{\boldsymbol u}_2=-\omega_3\frac{\partial}{\partial \omega_1} + \omega_1\frac{\partial}{\partial \omega_3}
-\gamma_3\frac{\partial}{\partial \gamma_1} + \gamma_1\frac{\partial}{\partial \gamma_3}
-R_3\frac{\partial}{\partial R_1} + R_1\frac{\partial}{\partial R_3}, \\
\hat{\boldsymbol u}_3=-\omega_2\frac{\partial}{\partial \omega_1} + \omega_1\frac{\partial}{\partial \omega_2}
-\gamma_2\frac{\partial}{\partial \gamma_1} + \gamma_1\frac{\partial}{\partial \gamma_2}
-R_2\frac{\partial}{\partial R_1} + R_1\frac{\partial}{\partial R_2}.
\end{eqnarray*}
A Poincar\'{e} map for this case is shown in Fig.~\ref{fig_Sym}$a$, and the motion of the point is illustrated in Fig.~\ref{fig_Sym}$b$.
As can be seen, the map is foliated by invariant curves, and the point of contact moves in a circle.
For explicit integration of the system (\ref{eq10}) in this case it is more convenient to pass to the fixed coordinate system
(see~\cite{JacobiIntegral} for details).

\begin{figure}[!ht]
	\begin{center}
		\includegraphics[totalheight=4.5cm]{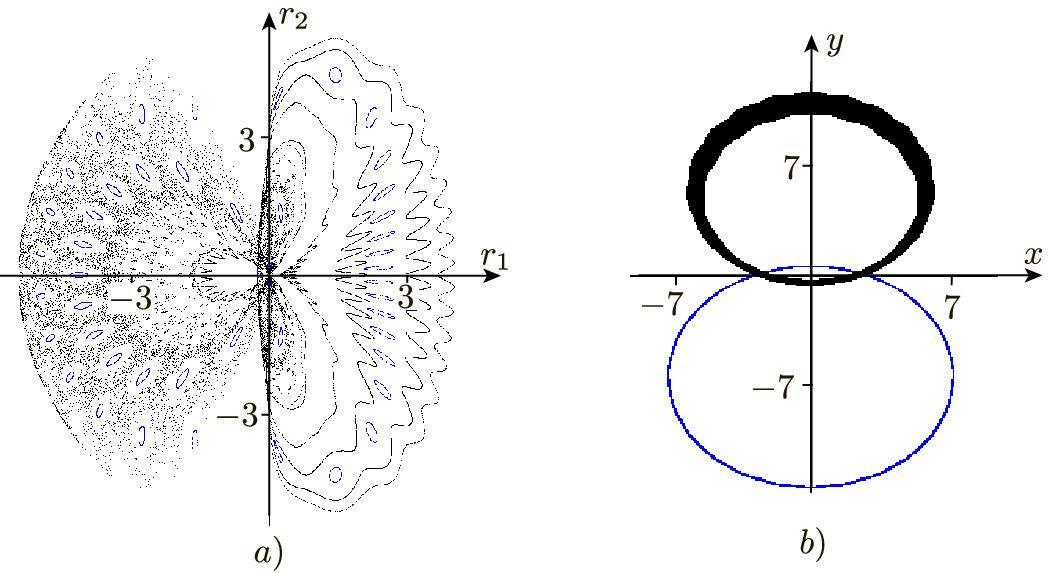}
		\caption{A Poincar\'{e} map for a dynamically symmetric ball ($a$) and the trajectory of the contact point ($b$)
			for fixed parameters $m=1$, $a=5$, $I_1=I_2=2$ $I_3=3$, $H=6$, $h=5$, $G=10$, $\Omega=2$.  }
		\label{fig_DS}
	\end{center}
\end{figure}

For a dynamically symmetric ball ($I_1=I_2$) the equations of motion (\ref{eq10}) possess only one symmetry field $\hat{\boldsymbol u}_3$.
In this case, the variable $l$ is cyclic (i.\,e., $\hat{\boldsymbol u}_3(l)=0$). Thereby the problem reduces to investigating
a two-dimensional Poincar\'{e} map.
As is evident from Fig.~\ref{fig_DS}$a$, the map has in this case a chaotic trajectory, and hence there is no additional integral.
This implies that the conclusion made by Tz\'{e}noff in~\cite{Tzenoff} concerning integrability of this case by quadratures is incorrect.

\begin{figure}[!ht]
	\begin{center}
		\includegraphics[totalheight=4.5cm]{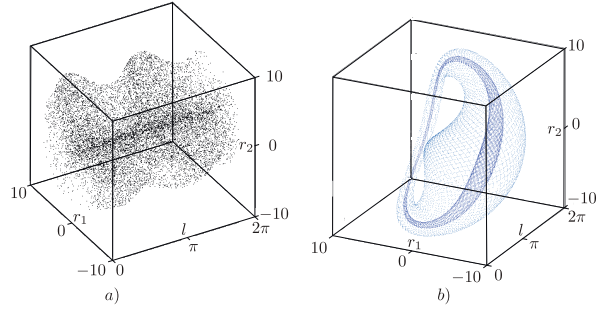}
		\caption{A Poincar\'{e} map for a dynamically asymmetric ball at different values $a)$ $H=2$ and $b)$ $H=6$,
			    the other parameters have the values $m=1$, $a=5$, $I_1=2$, $I_2=3$ $I_3=4$, $h=5$, $G=10$, $\Omega=2$.  }
		\label{fig_map3d}
	\end{center}
\end{figure}

\begin{figure}[!ht]
	\begin{center}
		\includegraphics[totalheight=4.5cm]{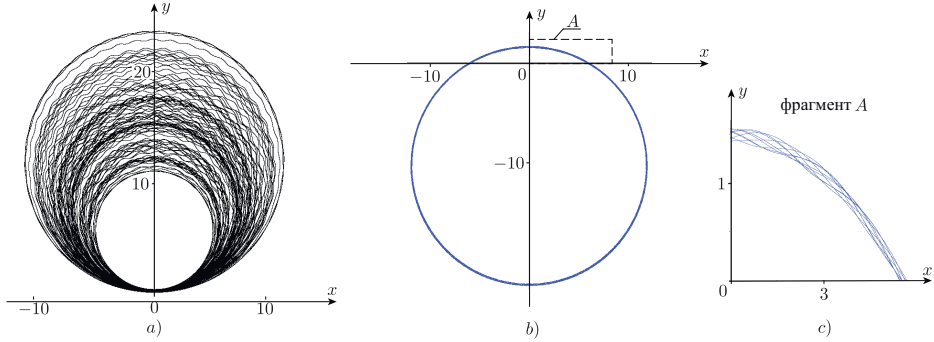}
		\caption{Trajectory of the contact point of a dynamically asymmetric ball for parameters corresponding to Fig.\,\ref{fig_map3d}.}
		\label{fig_CP}
	\end{center}
\end{figure}

A Poincar\'{e} map for a dynamically asymmetric ball is shown in Fig.~\ref{fig_map3d}. The presence of a chaotic trajectory in
Fig.~\ref{fig_map3d}$a$ shows the absence of two additional integrals. However, the map has invariant tori (see Fig.~\ref{fig_map3d}$b$).
Numerical experiments show that in both cases the trajectory of the contact point remains bounded (see Fig.~\ref{fig_CP}).

An interesting feature of the behavior of the contact point is that it is not only bounded, but also regular in a sense, at least visually.
Indeed, although the reduced system has chaotic behavior, the latter is
very difficult to detect from visual inspection of the trajectory of the contact point, which lies between two circles
and, at first sight, has quasi-periodic behavior. As a result, the above-mentioned chaotic behavior
is very difficult to detect experimentally. A similar phenomenon is observed in the behavior of the trajectory of the contact point
of the Chaplygin top~\cite{Kazakov} and rattlebacks~\cite{RattlebackUFN}, the reduced phase space of which contains strange attractors.
 In this connection we refer the reader to the work~\cite{Kazakov}, in which the authors actually doubt the widely spread belief
that a long-range weather forecast is impossible due to the presence of a Lorenz attractor in the simplified hydrodynamical model.
It turns out that, although some dynamical variables (such as velocities)
may have ``strongly chaotic'' behavior, experimentally measurable variables can behave rather regularly.
In particular, the trajectories in Fig.~\ref{fig_CP} are obtained from the reduced system (whose map is shown in Fig.~\ref{fig_map3d}$a$)
by an additional quadrature, which is seen to have a ``regularizing'' character.

In contrast to the fixed plane, where the integrability of the system allows a complete analysis of the contact point~\cite{Drift},
the reduced  system (\ref{eq10}) exhibits in this case chaotic trajectories, and so it does not seem possible to completely
describe the motion of the contact point. The problem of finite motion of the contact point is also unresolved.
It is possible that there exist unbounded trajectories due to the phenomenon of diffusion described for three-dimensional maps in the recent
paper~\cite{Simo}.


\section{The case $\boldsymbol{M}\parallel\boldsymbol{\gamma}$}
\label{sec5}

{\bf 1.} Let the angular momentum $\boldsymbol{M}$ be parallel to the normal vector $\boldsymbol{\gamma}$:
\begin{equation}
\label{Mpn}
\boldsymbol{M}=\lambda\boldsymbol{\gamma}, \quad \lambda= \tilde{F}_2={\rm const}.
\end{equation}
This case requires a separate analysis, since in this case the integrals $\tilde{F}_1$ and
$\tilde{F}_2$
are dependent.  If the plane is fixed ($\Omega=0$), the system is integrable
and, on a fixed level set of the energy integral, the evolution of the normal $\boldsymbol{\gamma}$
is governed by the Euler equation for the motion of a free rigid body with a fixed
point~\cite{Transformation,Chaplygin}.

In view of (\ref{Mpn}) the relation for the angular velocity has the form
\begin{eqnarray}
\label{sved}
\boldsymbol{\omega}={\bf A}\boldsymbol{K} + \frac{(\boldsymbol{\gamma}, {\bf A}\boldsymbol{K})}{d^{-1} - (\boldsymbol{\gamma},
{\bf A}\boldsymbol{\gamma})}{\bf A}\boldsymbol{\gamma}, \\
{\bf A}=diag(a_1, a_2, a_3),  \quad
\boldsymbol{K}=\lambda \boldsymbol{\gamma} + ma\Omega \boldsymbol{R}, \\
\quad a_i=(I_i + d)^{-1},  \quad d=ma^2.
\end{eqnarray}
As a result, we obtain a closed system of equations governing the evolution of $\boldsymbol{K}$ and
$\boldsymbol{\gamma}$ in the form
\begin{equation}
\label{Mpn2}
\dot{\boldsymbol{K}}=\Omega\boldsymbol{\gamma}\times\boldsymbol{K} + (\boldsymbol{K} - d\Omega\boldsymbol{\gamma})\times\boldsymbol{\omega},\\
\dot{\boldsymbol{\gamma}}=\boldsymbol{\gamma}\times\boldsymbol{\omega}.
\end{equation}

The first integrals of this system can be represented as
$$
\boldsymbol{\gamma}^2=1, \quad ( \boldsymbol{K},\boldsymbol{\gamma})=\lambda, \\
\tilde{E}=\frac{1}{2}(\boldsymbol{K}, {\bf A}\boldsymbol{K}) - \frac{\boldsymbol{K}^2}{2 d} + \frac{d}{2(1 - d(\boldsymbol{\gamma},
{\bf A}\boldsymbol{\gamma}))}({\bf A}\boldsymbol{K}, \boldsymbol{\gamma})^2.
$$
In addition, this system possesses the invariant measure
$$
\mu=\rho d\boldsymbol{K}d\boldsymbol{\gamma}, \quad \rho=\big(1 - d(\boldsymbol{\gamma}, {\bf A}\boldsymbol{\gamma})\big)^{-\frac{1}{2}}.
$$

We see that the first integrals and the invariant measure of the system (\ref{Mpn2}) are
analogous to the invariants in the Chaplygin problem of a ball rolling on a fixed
plane~\cite{Chaplygin, BM_gam}. This raises a natural question of the possibility of
representing the equations of motion (\ref{Mpn2}) in conformally Hamiltonian form
(see~\cite{BolsinovGeometrisation, Balseiro} for details).
This problem will not be considered here.

{\bf 2.} For the system (\ref{Mpn2}) to be integrable, we need an additional integral.
We show that in the general case it is absent. For this we investigate numerically
the Poincar\'{e} map.

We define the Andoyer\,--\,Deprit variables $(L,G,l,g)$
\begin{eqnarray*}
K_1=\sqrt{G^2 - L^2}\sin l, \quad K_2=\sqrt{G^2 - L^2}\cos l, \quad K_3=L, \\
\gamma_1=\left(\frac{\lambda}{G}\sqrt{1 - \frac{L^2}{G^2}} + \frac{L}{G}\sqrt{1 - \frac{\lambda^2}{G^2}} \right) \sin l + \sqrt{1 - \frac{\lambda^2}{G^2}}\sin g\cos l, \\
\gamma_2=\left(\frac{\lambda}{G}\sqrt{1 - \frac{L^2}{G^2}} + \frac{L}{G}\sqrt{1 - \frac{\lambda^2}{G^2}} \right) \cos l - \sqrt{1 - \frac{\lambda^2}{G^2}}\sin g\sin l, \\
\gamma_3=\lambda\frac{ L}{G^2} - \sqrt{1 - \frac{L^2}{G^2}}\sqrt{1 - \frac{\lambda^2}{G^2}}\cos g,
\end{eqnarray*}
where $l,g\in [0, 2\pi)$ and $L$, $G$ satisfy the inequality
$$
-1\leqslant\frac{L}{G}\leqslant1.
$$
On a fixed level set of the integral $\tilde{E}=h$ the system (\ref{Mpn2})
describes a four-dimensional flow. We choose $g=0$ as a secant of this flow and obtain a
two-dimensional Poincar\'{e} map,
which we parameterize by the variables $\left( l,\frac{L}{G}\right)$.
\begin{figure}[!ht]
	\begin{center}
		\includegraphics[totalheight=4.5cm]{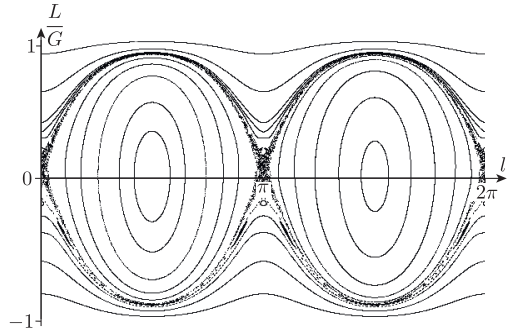}
		\caption{ A Poincar\'{e} map of the system (\ref{Mpn2}) for fixed parameters
			$a_1=1$, $a_2=2$, $a_3=3$, $d=4$, $h=100$, $\lambda=2$, $\Omega=0.5$.}
		\label{fig_map2d}
	\end{center}
\end{figure}

A typical view of the Poincar\'{e} map is presented in Fig.~\ref{fig_map2d}. In this case,
the map exhibits chaotic trajectories and hence, in the general case,
there is no additional integral.

{\bf 3.} If the case is dynamically symmetric ($a_1=a_2$), there exists an additional
integral $F_3$ (linear in $\boldsymbol{K}$):
$$
F_3=\rho\gamma_3(K_1\gamma_1 + K_2\gamma_2) - \rho\left(1 - \gamma_3^2 - \frac{1}{a_1d} \right)K_3 + \Omega \Psi(\gamma_3),
$$
where the function $\Psi(\gamma_3)$ is defined, depending on the moments of inertia, by the following
relations:
\begin{itemize}
	\item[] If $a_3>a_1$,
	 then
	 $$
	\Psi(\gamma_3)=\frac{ \sqrt{d}}{2\sqrt{a_3-a_1}}\left(1 - \frac{1}{a_1d} \right)^2\arctan
\left( \rho\frac{1 - a_1d - 2d(a_3-a_1)\gamma_3^2}{2\gamma_3\sqrt{d(a_3-a_1)}}\right). \quad
	$$
	\item[]  If $a_3<a_1$,
	then
	$$
	\Psi(\gamma_3)=\frac{ \sqrt{d}}{4\sqrt{a_1-a_3}}\left(1 - \frac{1}{a_1d} \right)^2\big(\ln(\xi + 1) - \ln(\xi - 1)  \big), \\
	\xi=\rho\frac{1 - a_1d - 2d(a_3-a_1)\gamma_3^2}{4\gamma_3\sqrt{d(a_1-a_3)}}.
	$$
\end{itemize}
Thus, in the case of a dynamically symmetric ball the system (\ref{Mpn2}) is integrable
by quadratures.

It can be shown that in the general case the fixed level set of the integral
$\tilde{E}(\boldsymbol{K}, \boldsymbol{\gamma})=h$ can, depending on the system parameters, define
a noncompact surface in phase space. However, numerical experiments show that the trajectories
of the system (\ref{Mpn2}) and the motion of the point of contact of the ball
are bounded. A rigorous analysis of the conditions of boundedness of the ball's
trajectories is a separate problem which can be solved by the methods of~\cite{BolsinovTopology}.

\newpage
\section{Discussion}
\label{sec6}

We point out some questions that are related to this work and require additional research.

For the motion of the Chaplygin ball on a rotating plane one can present the same generalizations of the problem that
were studied in detail for the Earnshaw problem of a homogeneous ball described in the Introduction. For example, it is possible to
introduce additional forces and friction torques, inclination of the plane and the action of gravity and to consider the now nonintegrable
scattering problem~\cite{ANAIS1}. However, all these generalizations are rather complicated and can evidently be
carried out only by numerical simulations. The study of these problems is of great importance for determining the scope of
applicability of the nonholonomic model and the role of friction which can lead to new interesting phenomena in natural experiments.

\section*{Acknowledgments}

The authors extend their gratitude to F.~Fass\'o and L.\,C.~Garc\'ia\,--\,Naranjo.

The work of A.\,V.~Borisov (Sections~\ref{sec1} and~\ref{sec2}) and I.\,A.~Bizyaev (Sections~\ref{sec3} and~\ref{sec4}) was carried out
within the framework of the state assignment of the Ministry of Education and Science of Russia (1.2404.2017/4.6).
The work of I.\,S.~Mamaev (Sections~\ref{sec5} and~\ref{sec6}) was carried out at MIPT under project 5-100 for state support for leading universities of the
Russian Federation.


\section*{References}
\begin{thebibliography}{99}
	

\bibitem{Alves}
Alves J and Dias J 2003
Design and Control of a~Spherical Mobile Robot
\textit{J.~Syst. Control Eng.} {\bf 217} 457--467

\bibitem{Balseiro}
Balseiro P and Garc\'{\i}a-Naranjo L C 2012
Gauge Transformations, Twisted Poisson Brackets and Hamiltonization of Nonholonomic Systems
\textit{Arch. Ration. Mech. Anal.} {\bf 205} 267--310

\bibitem{Acceleration}
Bizyaev I A, Borisov A V and Mamaev I S 2017
The Chaplygin Sleigh with Parametric Excitation: Chaotic Dynamics and Nonholonomic Acceleration
\textit{Regul. Chaotic Dyn.} {\bf 22} 955--975

\bibitem{BolsinovTopology}
Bolsinov A V, Borisov A V and Mamaev I S 2010
Topology and Stability of Integrable Systems
\textit{Russian Math. Surveys} {\bf 65} 259--318
see also:
\textit{Uspekhi Mat. Nauk} {\bf 65} 71--132

\bibitem{BolsinovGeometrisation}
Bolsinov A V, Borisov A V and Mamaev I S 2015
Geometrisation of Chaplygin's Reducing Multiplier Theorem
\textit{Nonlinearity} {\bf 28} 2307--2318

\bibitem{Kuznetsov}
Borisov A V, Jalnine A Yu, Kuznetsov S P, Sataev I R and Sedova J V 2012
Dynamical Phenomena Occurring due to Phase Volume Compression in Nonholonomic Model of the Rattleback
\textit{Regul. Chaotic Dyn.} {\bf 17} 512--532

\bibitem{Sataev}
Borisov A V, Kazakov A O and Sataev I R 2014
The Reversal and Chaotic Attractor in the Nonholonomic Model of Chaplygin's Top
\textit{Regul. Chaotic Dyn.} {\bf 19} 718--733

\bibitem{Kazakov}
Borisov A V, Kazakov A O and Pivovarova E N 2016
Regular and~Chaotic Dynamics in~the~Rubber Model of~a~Chaplygin Top,
\textit{Regul. Chaotic Dyn.} {\bf 21} 885--901

\bibitem{Transformation}
Borisov A V, Kilin A A and Mamaev I S 2012
Generalized Chaplygin's Transformation and Explicit Integration of a~System with a~Spherical Support
\textit{Regul. Chaotic Dyn.} {\bf 17} 170--190

\bibitem{Drift}
Borisov A V, Kilin A A and Mamaev I S 2013
The Problem of Drift and Recurrence for the Rolling Chaplygin Ball
\textit{Regul. Chaotic Dyn.} {\bf 18} 832--859

\bibitem{JacobiIntegral}
Borisov A V, Mamaev I S and Bizyaev I A 2015
The Jacobi Integral in Nonholonomic Mechanics
\textit{Regul. Chaotic Dyn.} {\bf 20} 383--400

\bibitem{Historical}
Borisov A V, Mamaev I S and Bizyaev I A 2016
Historical and~Critical Review of~the~Development of~Nonholonomic Mechanics: The~Classical Period
\textit{Regul. Chaotic Dyn.} {\bf 21} 455--476

\bibitem{UMN}
Borisov A V, Mamaev I S and Bizyaev I A 2017
Dynamical Systems with Non-Integrable Constraints: Vaconomic Mechanics, Sub-Riemannian Geometry, and Non-Holonomic Mechanics
\textit{Russian Math. Surveys} {bf 72} 783--840
see also:
\textit{Uspekhi Mat. Nauk} {\bf 72} 3--62

\bibitem{*2}
Borisov A V, Mamaev I S and Kilin A A 2002
Rolling of~a~Ball on~a~Surface: New Integrals and Hierarchy of~Dynamics
\textit{Regul. Chaotic Dyn.}{\bf 7} 201--219

\bibitem{Hierarchy}
Borisov A V and Mamaev I S 2002
The Rolling Motion of~a~Rigid Body on~a~Plane and~a~Sphere: Hierarchy of Dynamics
\textit{Regul. Chaotic Dyn.} {\bf 7} 177--200

\bibitem{RattlebackUFN}
Borisov A V and Mamaev I S 2003
Strange Attractors in Rattleback Dynamics
\textit{Physics–Uspekhi} {\bf 46} 393--403
see also:
\textit{Uspekhi Fiz. Nauk} {\bf 173} 407--418

\bibitem{BMF}
Borisov A V and Mamaev I S 2013
Topological Analysis of~an~Integrable System Related to~the~Rolling of~a~Ball on~a~Sphere
\textit{Regul. Chaotic Dyn.} {\bf 18} 356--371

\bibitem{BorMam2015}
Borisov A V and Mamaev I S 2015
Symmetries and Reduction in Nonholonomic Mechanics
\textit{Regul. Chaotic Dyn.} {\bf 20} 553--604

\bibitem{Conticelli}
Camicia C, Conticelli F and Bicch A 2000
Nonholonomic Kinematics and Dynamics of the Sphericle in
\textit{Proc. of the IEEE/RSJ International Conference on Intelligent Robots and Systems (IROS, Takamatsu, Japan)}
{\bf 1} 805--810

\bibitem{Cartwright}
Cartwright J H E, Feingold M and Piro O 1999
An Introduction to~Chaotic Advection
in
\textit{Mixing: Chaos and Turbulence}
H Chat\'{e}, E Villermaux end J-MChomaz (Eds.) 1999
NATO ASI Series (Series~B: Physics) {\bf 373}
(Boston,\,Mass.: Springer) 307--342

\bibitem{Chaplygin}
Chaplygin S A 2002
On a Ball's Rolling on a Horizontal Plane
\textit{Regul. Chaotic Dyn.} {\bf 7} 131--148
see also:
Chaplygin S A 1903
On a Ball's Rolling on a Horizontal Plane
\textit{Math. Sb.} {\bf 24} 139--168

\bibitem{Sun}
Cheng Ch Q and Sun Y S 1989/90
Existence of~Invariant Tori in~Three-Dimensional Measure-Preserving Mappings
\textit{Celest. Mech. Dynam. Astronom.} {\bf 47} 275--292

\bibitem{Dullin}
Dullin H R and Meiss J D 2009
Quadratic Volume-Preserving Maps: Invariant Circles and~Bifurcations,
\textit{SIAM J.~Appl. Dyn. Syst.} {\bf 8} 76--128

\bibitem{Er}
Earnshaw S 1844
\textit{Dynamics, or an~Elementary Treatise on~Motion}
3rd ed. (Cambridge: Deighton)


\bibitem{Fas}
Fass\`{o} F and Sansonetto N 2015
Conservation of Energy and Momenta in Nonholonomic Systems with Affine Constraints
\textit{Regul. Chaotic Dyn.} {\bf 20} 449--462

\bibitem{Fass}
Fass\`{o} F and Sansonetto N 2016
Conservation of 'Moving' Energy in Nonholonomic Systems with Affine Constraints and Integrability of Spheres on Rotating Surfaces
\textit{J.~Nonlinear Sci.} {\bf 26} 519--544

\bibitem{Naranjo}
Fass\`{o} F, Garc\'{\i}a-Naranjo L C and Sansonetto N 2018
Moving Energies As First Integrals of Nonholonomic Systems with Affine Constraints
\textit{Nonlinearity} {\bf 31} 755--782

\bibitem{bor3}
Gersten J, Soodak H and Tiersten M S 1992
Ball Moving on Stationary or Rotating Horizontal Surface
\textit{Am. J.~Phys.} {\bf 60} 43--47

\bibitem{ANAIS2}
Ivanov A P 2016
The ANAIS Billiard Experiment
\textit{Dokl. Phys.} {\bf 61} 285--287
2016, vol.\,61, no.\,6, pp.\,285--287;
see also:
\textit{Dokl. Akad. Nauk} {\bf 468} 401--402

\bibitem{James}
Mireles James J D 2013
Quadratic Volume-Preserving Maps: (Un)stable Manifolds, Hyperbolic Dynamics, and Vortex-Bubble Bifurcations
\textit{J.~Nonlinear Sci.} {\bf 23} 585--615

\bibitem{KozlovRev}
Kozlov V V 2016
The Phenomenon of~Reversal in~the~Euler\,--\,Poincar\'{e}\,--\,Suslov Nonholonomic Systems
\textit{J.~Dyn. Control Syst.} {\bf 22} 713--724

\bibitem{ANAIS1}
Levy-Leblond J-M 1986
The ANAIS Billiard Table
\textit{Eur. J.~Phys.} {\bf 7} 252--258

\bibitem{Simo}
Meiss J D, Miguel N, Sim\'o C and Vieiro A 2018
Accelerator Modes and~Anomalous Diffusion in~$3$D Volume-Preserving Maps
\textsf{arXiv:1802.10484}

\bibitem{Milne}
Milne E A 1948
\textit{Vectorial Mechanics}
(New York: Interscience)

\bibitem{Rays}
Routh E J 1955
\textit{The Advanced Part of~a~Treatise on~the Dynamics of~a~System of~Rigid Bodies:
Being Part~II of~a~Treatise on~the~Whole Subject}
6th ed. (New York: Dover)

\bibitem{Tzenoff}
Tz\'enoff I 1925
Quelques formes diff\'erentes des \'equations g\'en\'erales du mouvement des syst\'emes mat\'eriels
\textit{Bull. Soc. Math. France} {\bf 53} 80--105

\bibitem{Tzenoff2}
Tz\'enoff I 1920
Sur les \'equations g\'en\'erales du mouvement des syst\`emes mat\'eriels non holonomes
\textit{J.~Math. Pures Appl.~(8)} {\bf 3} 245--263

\bibitem{Fedorov}
Borisov A V, Fedorov Yu N 1995
On two modified integrable problems in dynamics
\textit{Mosc. Univ. Mech. Bull.} {\bf 50} 16--18
see also:
\textit{Vestnik Moskov. Univ. Ser.~1. Mat. Mekh.} {\bf 6} 102--105

\bibitem{BM_gam}
Borisov A V and Mamaev I S 2001
Chaplygin's Ball Rolling Problem Is Hamiltonian
\textit{Math. Notes} {\bf 70} 720--723
see also:
\textit{Mat. Zametki} {\bf 70} 793--795

\bibitem{KozlovUSM}
Kozlov V V 1985
On the Theory of Integration of the Equations of Nonholonomic Mechanics
\textit{Uspekhi Mekh.} {\bf 8} 85--107 (Russian)

\bibitem{NF}
Neimark Ju I and Fufaev N A 1972
\textit{Dynamics of~Nonholonomic Systems}
Trans. Math. Monogr. {\bf 33}
(Providence,\,R.I.: AMS)


\end{thebibliography}
\end{document}